\documentclass[letterpaper]{jpsj2}

\usepackage{bm}
\title{Exchange and Spin Jahn-Teller Distortions for Triangular Cluster of Spin-1/2}
\author{Ikumi \textsc{Honda} and Kiyosi \textsc{Terao}\thanks{E-mail address: terk005@shinshu-u.ac.jp}}
\inst{Graduate School of Science and Technology,\\
and Faculty of Science, Shinshu University,\\ 1-1 Asahi-3, Matsumoto 390-8621}

\recdate{December 25, 2005}

\abst{
We study the effects of magnetoelastic coupling on the degenerate ground state
of the spin-1/2 antiferromagnetic Heisenberg model for a regular triangular spin cluster. 
Static displacement of spins spontaneously lifts the degeneracy of the ground state 
through the distance dependence of exchange coupling, i.e., a spin Jahn-Teller mechanism takes place.
On the other hand, dynamical displacement does not lift the degeneracy, although the cluster distorts spontaneously.
The energy decrease obtained by dynamical theory is twice that obtained by static theory because of quantum fluctuation. }

\kword{spin Jahn-Teller effect, exchange striction, Heisenberg model, quantum spin}

\begin{document}
\maketitle

\section{Introduction} 
For the spin-1/2 antiferromagnetic (AF) Heisenberg Hamiltonian 
 on a regular triangular spin cluster, the degeneracy of the ground state is fourfold.
When exchange parameter $J$ depends on the distance between spins,
it is interesting to know whether the degeneracy is lifted by spontaneous distortion or not,
in other words, whether 
a kind of Jahn-Teller effect due to spin coupling takes place or not.
The spin-driven Jahn-Teller effect was studied for a tetrahedron cluster by Yamashita and Ueda~\cite{yam}$^)$ with quantum spins, 
by Tchernyshyov $et$ $al.$~\cite{TCHER,TCHER_}$^)$ with classical spins, 
and by Terao~\cite{Ter,Ter_}$^)$ with classical spins for the pyrochlore lattice.
These works were developed on the basis of static distortion.
In this paper, we propose a dynamical model and conclude that spontaneous distortion certainly takes place 
but the degeneracy is not lifted.
In other words, the distortion is not a Jahn-Teller effect but an exchange striction. 

\section{Regular Triangular Spin Cluster}

The AF Heisenberg Hamiltonian on a regular-triangular cluster is 
\begin{align}
	\mathcal{H}_0 &= -2J_0 (\bm{s}_1 \cdot \bm{s}_2 +\bm{s}_2 \cdot \bm{s}_3 +\bm{s}_3 \cdot \bm{s}_1 ),
\end{align}
where $J_0 <$0 and each $s$=1/2.
The Hamiltonian $\mathcal{H}_0$ is invariant under the symmetry operations of the point group  $\rm{D_{3h}}$ $(\bar6 \rm{m}2)$. We represent $2^3$ spin states in terms of $z$  components of $\bm{s}_\ell\ (\ell=1,2,3)$ as shown in Table \ref{SpinZyotai}, where they are classified by the $z$  component of total spin, $S_z$.
\begin{table}[tbhp]
\caption{Spin states of triangular spin cluster and $S_z$.}
\label{SpinZyotai}
\begin{tabular}{ccc}
\ \ $S_z$ & \\
 \hline
     $+3/2$    & $|+++>$  \\
     $+1/2$   & $|++->$\ \ $|+-+>$\ \ $|-++>$  \\
     $-1/2$   & $|+-->$\ \ $|-+->$\ \ $|--+>$  \\
     $-3/2$   & $|--->$  \\
\hline
\end{tabular}
\end{table}
The states of $S_z=\pm 3/2$ belong to the ${\rm{A'}}_1$ representation.
Each triple state of $S_z=\pm 1/2$ is reduced to a singlet ${\rm{A'}}_1$ and a doublet ${\rm{E'}}$.
The basis of representation is written as $| \Gamma i ,S_z>$,
where $\Gamma i$ stands for the $i$th component of the $\Gamma$ representation.
By operating $\mathcal{H}_0$ on the bases, we have
\begin{align}
	\mathcal{H}_0 |{\rm{A'}}_1,S_z> &=-\frac{3}{2}J_0 |{\rm{A'}}_1,S_z>
\end{align}
for $S=3/2$, and
\begin{align}
	\mathcal{H}_0 |{\rm{E'}}i,S_z> &=\frac{3}{2}J_0 |{\rm{E'}}i,S_z>, \ \  i=1,\ 2,
\end{align}
for $S=1/2$.
Because $J_0<0$, the ground state is the doublet ${\rm{E'}}$.

\section{Perturbation Hamiltonian}
\begin{figure}[hb]
\begin{center}
\includegraphics[width=7cm]{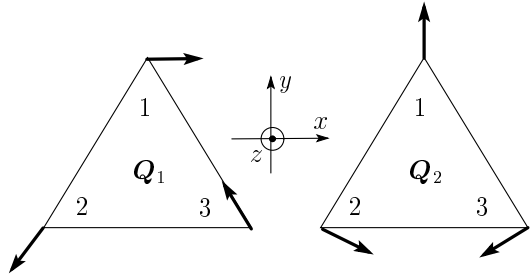}
\end{center}
\caption{Displacement caused by doublet $\rm{E'}$ mode.}
\label{E12}
\end{figure}

When $J$ depends on the distance between spins, the spin states are perturbed by distortion of the cluster.
To describe elastic vibrations around the stationary positions $\bm{R}^{0}_{\ell}$ of spins, we denote small deviations by $\bm{R}_\ell=\bm{R}_\ell ^0 +\bm{u}_\ell,(\ell=1,2,3),$
and $J$ is assumed to be a function of spin distance as $J(| \bm{R}_\ell -\bm{R}_{\ell'} | )$.
Normal coordinates of the cluster are determined by reducing the nine-dimensional representation ${\mathsf{D}}$ 
of point group ${\rm{D}_{\rm{3h}}}$ by $\bm{u}_{i}$'s as
\begin{align}
\mathsf{D} &=\rm{A_1'}+\rm{A_2'}+\rm{A_2''}+\rm{E''}+2\rm{E'}.
\end{align}
The normal vibrational modes are classified into the singlet $\rm{A_1'}$ ($Q_{\rm{A}}$: its normal coordinate) and doublet $\rm{E'}$ ($Q_1,\ Q_2$) representations
after eliminating the uniform translations ($\rm{A_2''}$,\ $\rm{E'}$) and uniform rotations ($\rm{A_2'}$,\ $\rm{E''}$).
The normal modes are represented by nine-dimensional vectors $(\bm{u}_{1};\bm{u}_{2};\bm{u}_3)$ as follows:
\begin{align}
	\bm{Q}_{\rm{A}} &=\frac{Q_{\rm{A}}}{\sqrt3}
	\Bigl(   0,1,0\ ;-\frac{\sqrt{3}}{2},-\frac{1}{2},0\ ;\frac{\sqrt{3}}{2},-\frac{1}{2},0 
	\Bigr),
\end{align}
\begin{subequations}
\begin{align}
	\bm{Q}_1&=\frac{Q_1}{\sqrt3}
		\Bigl(1,0,0 \ ;-\frac{1}{2},-\frac{\sqrt3}{2},0\ ;-\frac{1}{2},\frac{\sqrt3}{2},0
		\Bigr), \\
	\bm{Q}_2&=\frac{Q_2}{\sqrt3}
		\Bigl(0,1,0 \ ;\frac{\sqrt{3}}{2},-\frac{1}{2},0\ ;-\frac{\sqrt{3}}{2},-\frac{1}{2},0
		\Bigr).
\end{align}
\end{subequations}
The singlet $Q_{\rm{A}}$ is the breathing mode.
The doublet E$'$ ($Q_1,\,Q_2$) is illustrated in Fig.
 \ref{E12}.
The bases of the irreducible representations made from the bilinear combinations of the spin operators for
the singlet ${\rm{A}^{\prime }_1}$ ($f_{\rm{A}}$) and the doublet $\rm{E}^{\prime }$ ($f_1,\ f_2$) are
\begin{align}
	f_{\rm{A}} 
	&=\frac{1}{\sqrt3}(\bm{s}_1\cdot\bm{s}_2+\bm{s}_2\cdot\bm{s}_3+\bm{s}_3\cdot\bm{s}_1), 
			\label{eqkanaNO} 
\end{align}
and
\begin{subequations}
\begin{align}
	f_1 &=\displaystyle \frac{1}{\sqrt2}
                       (\bm{s}_1\cdot\bm{s}_2-\bm{s}_3\cdot\bm{s}_1),
			\label{eqkanaHI}\\
	f_2 &=\displaystyle \frac{1}{\sqrt6}
                       (\bm{s}_1\cdot\bm{s}_2-2\bm{s}_2\cdot\bm{s}_3+\bm{s}_3\cdot\bm{s}_1).
			\label{eqkanaO}
\end{align}
\end{subequations}
 
The perturbation Hamiltonian is the sum of the elastic vibrations and the coupling between $Q_\alpha $'s and $f_\alpha$'s, i.e.,
\begin{align}
\mathcal{H'} = & \frac{1}{2m}({P_{\rm{A}}}^2+{P_1}^2+{P_2}^2) \nonumber\\
                  & +\frac{ m {\omega_{\rm{A}}}^2}{2} {Q_{\rm{A}}}^2
                     +\frac{ m {\omega_{\rm{E}}}^2}{2} ( {Q_1}^2 + {Q_2}^2) \nonumber\\
                  & -\{ J'_{\rm{A}}Q_{\rm{A}}f_{\rm {A}}+J'_{\rm{E}}(Q_1 f_1+ Q_2 f_2) \},
			\label{eqkanaNU}
\end{align}
where $J'_\alpha $'s are the coupling constants due to a change in $J$ caused by $Q_\alpha $ modes and $m$ is the mass of ions. When $J$ depends only on the distance between a spin pair, 
\begin{align}
J'_{\rm{A}} & =2\sqrt3
 \left( \frac{\partial J}{\partial r}
 \right), 
             & J'_{\rm{E}}=\sqrt6
 \left(  \frac{\partial J}{\partial r}
 \right).
\end{align}
By operating $f_\alpha $ on the singlet state $| {\rm{A}^{\prime }_1},S_z>$, we have
\begin{align}
&f_{\rm{A}} | {\rm{A}^{\prime }_1},S_z>=\frac{\sqrt3}{4}| {\rm{A}^{\prime }_1},S_z >, \\
&f_1 | {\rm{A}^{\prime }_1},S_z > =f_2 | {\rm{A}^{\prime }_1},S_z >=0. 
\end{align}
In the subspace of the doublet state $|{\rm{E'}}i,S_z>$,
\begin{align}
	& f_{\rm{A}} =-\frac{\sqrt3}{4} \sigma_1,\\
	& f_1 =\frac{\sqrt6}{4}\sigma_x,\ \ 	f_2 = \frac{\sqrt6}{4}\sigma_z,
	\label{eqkanaA}
\end{align}
using the $2\times2$ unit matrix $\sigma_1$, and the Pauli matrices $\sigma _x$ and $\sigma _z$. 

\section{Static Theory}
In the case of static displacement of spins, the perturbation $\mathcal{H'}$, 
eq. (\ref{eqkanaNU}), is rewritten in the subspace of the $\rm{E}^{\prime }$ 
states as
\begin{align}
	\mathcal{H}^{\prime} 
  &  =\frac{ m{\omega_{\rm{A}}}^2 }{ 2 }
        {Q_{\rm{A}}}^2+\frac{m{\omega_{\rm{E}}}^2}{2}({Q_1}^2+{Q_2}^2 )\nonumber\\
  &  +\frac{ \sqrt3{J'_{\rm{A}}} }{ 4 }
        Q_{\rm{A}}
      -\frac{ \sqrt6{J'_{\rm{E}}} }{ 4 }
       (Q_1 \sigma _x +Q_2 \sigma _z).
\end{align}
By diagonalizing this, we obtain eigenvalues as
\begin{align}
	\delta E' & =\frac{ m{\omega_{\rm{A}}}^2 }{ 2 }  {Q_{\rm{A}}}^2
              +\frac{ m {\omega_{\rm{E}}}^2 }{ 2 }  ({Q_1}^2+{Q_2}^2 )\nonumber\\
           & +\frac{ \sqrt3{J'_{\rm{A}}} }{ 4 } 
                      Q_{\rm{A}} \mp \frac{\sqrt6 }{4}J'_{\rm{E}}\sqrt{{Q_1}^2+{Q_2}^2}.
\end{align}
The degeneracy of the ground state is lifted.
Minimizing $\delta E'$ with respect to variations in ${Q_{\rm{A}}}^2$ and ${Q_1}^2+{Q_2}^2(=\rho^2)$, we have
\begin{align}
	\delta E'_{\rm{min}} 
	&=- \frac{3}{ 32 m {\omega_{\rm{A}}}^2} {J'_{\rm{A}}}^2
	   - \frac{3}{16 m {\omega_{\rm{E}}}^2} {J'_{\rm{E}}}^2,
\end{align}
and
\begin{align}
	{Q_{\rm{A}}}^2 &=\frac{  3 {J'_{\rm{A}}}^2  }{ 16 m^2 {\omega_{\rm{A}}}^4 },
                 &\rho^2 =\frac{ 3{J'_{\rm{E}}}^2 }{ 8 m^2 {\omega_{\rm{E}}}^4 }.
			\label{eqkanaTE}%
\end{align}
Because $\delta E'_{\rm{min}}<0$, 
the spin-driven Jahn-Teller effect takes place as shown for
a tetrahedron by Yamashita and Ueda.~\cite{yam}$^)$

\section{Dynamical Theory}
\subsection{Expansion up to first order of distortion}
Now we treat dynamically the displacement of spins.
Creation and annihilation operators $b^{\dagger}_\alpha$ and $b_\alpha$ for normal mode $Q_\alpha $ are  
\begin{subequations}
\begin{align}
	& {b_\alpha}^\dagger = \frac{1}{\sqrt{2\hbar}}
	\Bigl(
		\sqrt{m\omega_\alpha }Q_\alpha -i\frac{1}{\sqrt{m\omega_\alpha }}P_\alpha
	\Bigr), \label{eqkanaYU}\\
	& b_\alpha = \frac{1}{\sqrt{2\hbar}}
	\Bigl(
		\sqrt{m\omega_\alpha }Q_\alpha+i\frac{1}{\sqrt{m\omega_\alpha }}P_\alpha
	\Bigr), 
			\label{eqkanaRI}
\end{align}
\end{subequations}
where $\alpha={\rm{A}},1,2$ ,
$\omega_{\rm{E}} =\omega_1=\omega_2$ and $J'_{\rm{E}}=J'_1=J'_2$. 
$\mathcal{H'}$, eq. (\ref{eqkanaNU}), is rewritten as
\begin{align}
	\mathcal{H'}= \sum_{\alpha } 
	\Bigl[
	\hbar\omega_\alpha (b^{\dagger}_\alpha b_\alpha +\frac{1}{2})
	-\sqrt{\frac{\hbar}{2m\omega_\alpha }} J'_\alpha f_\alpha (b_\alpha +b^\dagger_{\alpha} ) 
	\Bigr].
\end{align}
By introducing modified operators $\tilde b^{\dagger}_\alpha$ and $\tilde b_\alpha $,
\begin{subequations}
\begin{align}
	{\tilde{b}_\alpha}\,^{\dagger} = \ &{b_\alpha}^{\dagger}
	-\frac{J'_\alpha }{\hbar\omega_\alpha }\sqrt{\frac{\hbar}{2m\omega_\alpha }} f_\alpha,
     \ \ \ \ \ \   \\
	\tilde b_\alpha = \ & b_\alpha
	-\frac{J'_\alpha }{\hbar\omega_\alpha }\sqrt{\frac{\hbar}{2m\omega_\alpha }} f_\alpha,
	\ \ \ \ \ \ 
\end{align}
\end{subequations}
we have
\begin{align}
	\mathcal{H'} & =\sum_{\alpha }
	\Bigl[	\hbar\omega_\alpha
		\Bigl(   \tilde b^\dagger_\alpha \tilde b_\alpha +\frac{1}{2} \Bigr)
		-\frac{1}{2m {\omega_\alpha }^2 } {{J'}_{\alpha }}^2 f_{\alpha}^2
	\Bigr].
\end{align}
Because commutation relations of $\tilde b^{\dagger}_\alpha$ and $\tilde b_\alpha $ are
\begin{align}
[\tilde b_\alpha ,\tilde b_\alpha^{\dagger} ] 
&=1, 
&[\tilde b_\alpha ,\tilde b_\alpha ] =[\tilde b_\alpha^{\dagger} ,\tilde b_\alpha^{\dagger} ]=0,
\end{align}
$\tilde b_\alpha$ and $\tilde b_\alpha^{\dagger}$ are boson operators.
Commutation relations of $f_{\rm{A}}$ with $f_\alpha$ for $\alpha ={\rm{A}},1,2$ are
\begin{align}
 [f_{\rm{A}},f_\alpha ] =0.
\end{align}
With respect to $f_1$ and $f_2$, we have
\begin{align}
[f_1,f_2] &=\sqrt3 i \bm{s}_3 \cdot (\bm{s} _1 \times \bm{s} _2).
\end{align}
Then, the commutators for the E$'$ mode are
\begin{align}
	[\tilde b_1,\tilde b_2^{\dagger}]= [\tilde b_1,\tilde b_2]
	=[\tilde b_1^{\dagger},\tilde b_2^{\dagger}] \nonumber\\
	=\sqrt3 i \bm{s}_3 \cdot (\bm{s} _1 \times \bm{s} _2)
	\frac{{J'_{\rm E}}^2 }{2m\hbar{\omega_{\rm E}}^3 },
\label{eqkanaKO}
\end{align}
Although these commutation relations give rise to complexity in excited states, the ground state with respect to modified bosons is defined as 
\begin{align}
	\tilde b_\alpha  & |{\rm{E'}}i,S_z>_0 =0,
			\label{eqTeigiSiki}
\end{align}
or
\begin{align}
b_\alpha |{\rm{E'}}i,S_z>_0 
  & =\frac{J'_\alpha }{\hbar\omega_\alpha }\sqrt{\frac{\hbar}{2m\omega_\alpha }} f_\alpha |{\rm{E'}}i,S_z>_0,
\label{eqTeigiSiki_}
\end{align}
where subscript $0$ denotes the subspace of the ground states of $\tilde b_\alpha $.
In this subspace,
\begin{align}
	\mathcal{H'} & |{\rm{E'}}i,S_z>_0 \nonumber\\
	&=\sum_{\alpha}
	\left(
		-\frac{1}{2 {m\omega_\alpha }^2 } {{J'}_{\alpha }}^2 {f_{\alpha}}^2
		+\frac{\hbar\omega_\alpha }{2}
	\right) |{\rm{E'}}i,S_z>_0 .
\end{align}
Because
\begin{align}
	{f_{\rm{A}}}^2 |{\rm{E'}}i,S_z>_0 &=\frac{3}{16} |{\rm{E'}}i,S_z>_0,
\end{align}
and
\begin{align}
	{f_1}^2 |{\rm{E'}}i,S_z >_0 &={f_2}^2 |{\rm{E'}}i,S_z>_0 =\frac{3}{8} |{\rm{E'}}i,S_z>_0
\end{align}
 for $i$=1, 2, the perturbation energy $\delta E'$ is obtained from the eigenvalue of $\mathcal{H'}$ as follows
\begin{align}
	\delta E' &=-\frac{3}{32 m \omega_{\rm{A}}^2}{ J'_{\rm{A} }}^2-\frac{3}{8m \omega_{\rm{E}}^2} {J'_{\rm{E}}}^2,
\end{align}
after eliminating the zero-point energy.
Thus, the distortion reduces the energy but does not lift the degeneracy of the ground spin state.
The $\delta E'$ due to the ${\rm{A'}}_1$ distortion in dynamical theory is equal to that in the static one, whereas the
$\delta E'$ due to the ${\rm{E'}}$ distortion in dynamical theory is twice that obtained in static theory.

The expected value of $Q_\alpha $ is  
\begin{align}
< {\rm{E'}}i,S_z & | Q_\alpha |{\rm{E'}}i,S_z>_0 \nonumber\\ 
&=< {\rm{E'}}i,S_z | \sqrt{\frac{\hbar}{2\omega}} (b_\alpha + b_\alpha^\dagger) |{\rm{E'}}i,S_z>_0,\\
&=< {\rm{E'}}i,S_z | \frac{1}{m\omega^2_\alpha  }J'_\alpha f_\alpha |{\rm{E'}}i,S_z>_0,
\end{align}
using eqs. (\ref{eqkanaYU}) and (\ref{eqkanaRI}). 
In the two-dimensional subspace of $|{\rm{E'}}i,S_z>$'s, $Q_{\alpha}$'s are represented as
\begin{align}
	& Q_{\rm{A}} = -\frac{\sqrt3}{4m{\omega _{\rm{A}}}^2}  J'_{\rm{A}} \sigma_1,
 \label{eqkanaKE} \\
	 & Q_1 =\frac{\sqrt6}{4m{{\omega_{\rm{E}}}^2 } } J'_{\rm{E}}\sigma_x, \hspace{1em}
	Q_2 = \frac{\sqrt6}{4m{{\omega_{\rm{E}}}}^2  } J'_{\rm{E}}\sigma_z.
\label{eqkanaRA}
\end{align}
The expected value of $Q_1$ is 
\begin{align}
	< {\rm{E'}}i, S_z | Q_1|{\rm{E'}}i,S_z > _0=0,
\end{align}
and the expected values of ${Q_\alpha} ^2$ with $\alpha =1,\ 2$ are  
\begin{align}
	<{\rm{E'}}i, S_z  | {Q_\alpha} ^2 |{\rm{E'}}i,S_z >_0  
	=\frac{3}{8m^2 {\omega_{\rm{E}}}^4 }  {J'_{\rm{E}}}^2+\frac{\hbar}{2m\omega_{\rm{E}} } ,
\end{align}
for $i$=1, 2.
Note that $Q_1$ fluctuates in contrast to $Q_{\rm{A}}$ and $Q_2$
as seen from eqs. (\ref{eqkanaKE}) and (\ref{eqkanaRA}). 
In appearance, the distortion caused by $Q_1$ is smeared out by the quantum fluctuation.
When $J'_{\rm{E}}>0$, the regular triangle is distorted by $Q_2$ 
into an acute isosceles triangle in the $|{\rm{E'}}1,S_z >$ state,
and into an obtuse one in the $|{\rm{E'}}2,S_z >$ state.
Although the triangle distorts into different shapes according to spin states, the degeneracy of the E$'$ spin state is not lifted.
The expected value of $\rho^2(= {Q_1}^2+ {Q_2}^2)$ is
\begin{align} 
	<{\rm{E'}}i,S_z | \rho^2 |{\rm{E'}}i,S_z >_0
	=\frac{3}{4m^2 \omega_{\rm{E}}^4}{J'_{\rm{E}}}^2+\frac{\hbar}{m\omega_{\rm{E}} }.
			\label{eqkanaE} 
\end{align} 
On the other hand, the sum of the expected values is
\begin{align}
{<{\rm{E'}}i,S_z  | Q_1|{\rm{E'}}i,S_z>_0}^2+&{<{\rm{E'}}i,S_z  | Q_{2}|{\rm{E'}}i,S_z>_0}^2\nonumber\\
&~~~~~~~=\frac{3{J'_{\rm{E}}}^2}{8m^2 \omega_{\rm{E}}^4},
\end{align}
which is half of eq. (\ref{eqkanaE}) and equal to the $\rho^2$ of eq. (\ref{eqkanaTE}). 
The energy change due to the ${\rm{E'}}$ distortion obtained by dynamical theory is twice that by static theory 
because of quantum fluctuation. 
\subsection{Expansion up to second order of distortion}
The bases of the irreducible representation of ${\rm{D}_{\rm{3h}}}$ composed of quadratic forms of $Q_\alpha $'s
are given for ${\rm{A'}_1}$ as
\begin{align}
	{Q_{\rm{A}}}^2,\ \frac{1}{\sqrt2}({Q_1}^2 +{Q_2}^2 ),
		\label{eqkanaU}
\end{align}
and for $\rm{E}^{\prime }$ as
\begin{subequations}
\begin{align}
	&{\rm E}'_1 :\ {Q_{\rm{A}}}{Q_1},\ \frac{1}{\sqrt2} (Q_1 Q_2 +Q_2 Q_1),
	\label{eqkanaSA}\\
	&{\rm E}'_2 :\ {Q_{\rm A}}{Q_2},\  \frac{1}{\sqrt2}({Q_1}^2 -{Q_2}^2).
	\label{eqkanaWI}
\end{align} 
\end{subequations}
The perturbation Hamiltonian is formulated from the couplings of eq. (\ref{eqkanaU}) with eq. (\ref{eqkanaNO}) for the ${\rm{A'}_1}$ symmetry,  and 
eqs. (\ref{eqkanaSA}) and (\ref{eqkanaWI}) with eqs. (\ref{eqkanaHI}) and (\ref{eqkanaO}) for the ${\rm{E'}}$ symmetry,
\begin{align}
\mathcal{H''}[{\rm{A'}_1}] &=-\Bigl[J''_{\rm{A}}{Q_{\rm{A}}}^2+\frac{J''_{\rm{AE}}}{\sqrt2}({Q_1}^2 +{Q_2}^2 )\Bigr]f_{\rm{A}}, \\
\mathcal{H''}[{\rm{E'}}] &=-\Bigl[J''_{\rm{EA}}Q_{\rm{A}}Q_1 +\frac{J''_{\rm{E}}}{\sqrt2} (Q_1 Q_2 +Q_2 Q_1)\Bigr] f_1 \nonumber\\ 
&+\Bigl[J''_{\rm{EA}}Q_{\rm{A}} Q_2 +\frac{J''_{\rm{E}}}{\sqrt2}({Q_1}^2-{Q_2}^2)\Bigr]f_2 .
\end{align}
When $J$ depends only on the distance between a spin pair, $J''_{\alpha\beta } $ are
\begin{align}
J''_{\rm{AA}} &=\frac{\sqrt3}{2}
	\Bigl( \frac{\partial^2 J}{\partial r^2}\Bigr),\\
J''_{\rm{AE}} &=\frac{\sqrt3}{4}
	\Bigl(\frac{1}{a }\frac{\partial J}{\partial r} +\frac{\partial^2 J}{\partial r^2} \Bigr),\\
J''_{\rm{EE}} &=\frac{1}{4}\sqrt{\frac{3}{2}}
	\Bigl(- \frac{1}{a}\frac{\partial J}{\partial r} +\frac{\partial^2 J}{\partial r^2} \Bigr),\\
J''_{\rm{EA}} &=\sqrt{\frac{3}{2}}
	\Bigl( \frac{\partial^2 J}{\partial r^2}\Bigr ),
\end{align}
where $a=| \bm{R}_\ell^0 -\bm{R}_{\ell'}^0 |$.
Now, the total Hamiltonian is
\begin{align}
	\mathcal{H} &=\mathcal{H}_0+\mathcal{H'}
	+\mathcal{H''}[{\rm{A'}_1}] +\mathcal{H''}[{\rm{E'}}].
\end{align}
In the subspace of the doublet $|{\rm E'}i, S_z> _0$ spin states,
using eq. (\ref{eqTeigiSiki_}), we have
\begin{align}
&\mathcal{H}''[{\rm A'}_1] +\mathcal{H}''[{\rm E'}]  \nonumber \\
=& \Bigl[ 
	\frac{J''_{\rm A}}{\omega_{\rm A}}
		\Bigl( \frac{ {J'_{\rm E} }^2}{m{\omega_{\rm A}}^3}
			{f_{\rm A}}^2 +\frac{\hbar}{2} \Bigr) \nonumber\\
		&+\frac{J''_{\rm AE}}{\sqrt2 \omega_{\rm E}}
			\Bigl(\frac{{J'_{\rm E}}^2}{m{\omega_{\rm E}}^3}({f_1}^2 +{f_2}^2 )
				+\frac{\hbar}{2}	\Bigr) f_{\rm A} \nonumber \\
		&+\frac{J''_{\rm EA} {J'_{\rm E}}^2}{{m(\omega_{\rm A}\omega_{\rm E})}^2}
			f_{\rm A}({f_1}^2 + {f_2}^2) \nonumber \\ 
		&+\frac{ J''_{\rm E} {J'_{\rm E}}^2}{\sqrt2 m {\omega_{\rm E}}^4}
			\{ (f_1 f_2 +f_2 f_1 ) f_1+({f_1}^2 -{f_2}^2 )f_2 \}
\Bigr],
\end{align}
where in the last term $f_1 f_2 + f_2 f_1$ and ${f_1}^2 -{f_2}^2$ vanish  and the other terms are proportional to the unit matrix because of eq. (\ref{eqkanaA}). 
The degeneracy in the spin ground state is not lifted. The energy correction obtained by the quadratic expansion is
\begin{align}
	\delta E'' & = -\frac{\sqrt{3}}{4}
	\Bigl[ 
		\frac{J''_{\rm{A}}}{\omega_{\rm{A}}} 
		\Bigl(
			\frac{3}{16}\frac{ {J'_{\rm{E}}}^2}{m{\omega_{\rm{A}}}^3}+\frac{\hbar}{2}
		\Bigr)	  \nonumber\\
	& +\frac{J''_{\rm{AE}}}{\sqrt2 \omega_{\rm{E}}}
	\Bigl(
		\frac{3}{4}\frac{ {J'_{\rm{E}}}^2 }{m{\omega_{\rm{E}}}^3}+\frac{\hbar}{2}
	\Bigr) 
	+\frac{3J''_{\rm{EA}}}{4m}{
		\Bigl(
			\frac{J'_{\rm{E}}}{\omega_{\rm{A}} \omega_{\rm{E }}}
		\Bigr) ^2}
	\Bigr]
\end{align}
for both components of the ground spin state of doublet $|{\rm E'}i, S_z> _0$.

\section{Conclusions and Discussion}
We have studied the frustrating quantum spin-1/2 system on the regular triangle.
When the exchange parameter depends on the distance between a spin pair,
the degeneracy is lifted in the static theory within the first order of distortion.
On the other hand,  
the degeneracy is not lifted in the dynamical theory, although spontaneous distortion takes place.
The change in energy obtained by dynamical theory is twice that obtained by the static theory because of the quantum fluctuation of $Q_1$ in E$'$ mode distortion.
Because the modified boson operators  for $\rm{E'}$ mode do not commute, as shown in eq. (\ref{eqkanaKO}),
excited states are not represented as the usual boson system.
It is an open question how the excited states are described.

\end{document}